\documentclass[letterpaper, 10pt, oneside, conference, final]{IEEEtran}
\usepackage[latin1]{inputenc}
\usepackage[pdftex]{graphicx}
\usepackage{array, multirow}
\usepackage{color}
\usepackage{booktabs}
\usepackage{caption}
\usepackage{subcaption}
\usepackage{tikz, pgfplots}
\usetikzlibrary{arrows.meta}
\usetikzlibrary{spy}
\usepackage{amssymb}
\usepackage{amsmath}
\usepackage{tabularx}
\usepackage{makecell}
\usepackage[top=0.742in, bottom=1.086in, left=0.68in, right=0.68in]{geometry}
\pgfplotsset{compat=1.16}
\def\BibTeX{{\rm B\kern-.05em{\sc i\kern-.025em b}\kern-.08em
		T\kern-.1667em\lower.7ex\hbox{E}\kern-.125emX}}

\usepackage[nolist]{acronym}
\begin{acronym}   
	\acro{DL}{deep learning}    
	\acro{DNN}{deep neural network} 
	\acro{LS}{least squares}   
	\acro{MLE}{maximum likelihood estimator}
	\acro{MSE}{mean squared error}
	\acro{RMSE}{root mean squared error}
	\acro{WSN}{wireless sensor network}
	\acro{RSS}{received signal strength}
	\acro{IoT}{Internet of Things}
	\acro{5G}{fifth generation}
	\acro{SU}{sensing unit}
	\acro{CU}{central unit}
	\acro{CRLB}{Cram\'{e}r-Rao lower bound}
	\acro{MPE}{mean position error}
	\acro{CDF}{cumulative distribution function}
	\acro{ToA}{time of arrival}
	\acro{AoA}{angle of arrival}
	\acro{REML}{radio environment map localization}
	\acro{OWL}{online wireless lab}
	\acro{AGV}{automated guided vehicle}
	\acro{SDR}{software-defined radio}
	\acro{OWL}{Online Wireless Lab}
	\acro{PLM}{path loss model}
	\acro{ELU}{exponential linear unit}
	\acro{MTL}{multi transmitter localization}
	\acro{CSI}{channel state information}
	\acro{CNN}{convolutional neural network}
	\acro{ReLU}{rectified linear unit}
	\acro{RG}{random guess}
	\acro{PS}{particle simulation}
\end{acronym}

\begin{document}
\IEEEoverridecommandlockouts
\title{Blind Transmitter Localization Using Deep Learning: A Scalability Study}
\author{
	\IEEEauthorblockN{Ivo Bizon\IEEEauthorrefmark{1}, Ahmad Nimr\IEEEauthorrefmark{1}, Philipp Schulz\IEEEauthorrefmark{1}, Marwa Chafii\IEEEauthorrefmark{2}\IEEEauthorrefmark{3} and Gerhard P. Fettweis\IEEEauthorrefmark{1}}
	\IEEEauthorblockA{\IEEEauthorrefmark{1}Vodafone Chair Mobile Communications Systems, Technische Universit{\"a}t Dresden (TUD), Germany}
	\IEEEauthorblockA{\{ivo.bizon, ahmad.nimr, philipp.schulz, gerhard.fettweis\}@ifn.et.tu-dresden.de}
	\IEEEauthorblockA{\IEEEauthorrefmark{2}Engineering Division, New York University (NYU), Abu Dhabi, UAE}
	\IEEEauthorblockA{\IEEEauthorrefmark{3}NYU WIRELESS, NYU Tandon School of Engineering, New York, USA}
	\IEEEauthorblockA{marwa.chafii@nyu.edu}
}
\maketitle
\begin{abstract}
	This work presents an investigation on the scalability of a deep leaning (DL)-based blind transmitter positioning system for addressing the multi transmitter localization (MLT) problem.
	The proposed approach is able to estimate relative coordinates of non-cooperative active transmitters based solely on received signal strength measurements collected by a wireless sensor network. 
	A performance comparison with two other solutions of the MLT problem are presented for demonstrating the benefits with respect to scalability of the DL approach.
	Our investigation aims at highlighting the potential of DL to be a key technique that is able to provide a low complexity, accurate and reliable transmitter positioning service for improving future wireless communications systems.
\end{abstract}
\begin{IEEEkeywords}
	Multi transmitter localization, network-side localization, wireless sensor network, received signal strength, deep learning, positioning.
\end{IEEEkeywords}
\acresetall

\section{Introduction}
\IEEEPARstart{I}{n} this paper a \ac{DL}-based framework that addresses the problem of estimating the relative position\footnote{The terms \emph{location} and \emph{position} are used interchangeably in this paper, and they refer to the Cartesian coordinates of one or more active transmitters within an area of interest, i.e., relative coordinates.} of multiple active transmitters within an area of interest using \ac{RSS} measurements collected by a \ac{WSN} as sole position information source is presented.
We focus on simultaneous blind transmitter localization, which refers to a functionality that can be implemented at the network side for enabling simultaneous localization of multiple transmitters  without prior assumption on the transmission protocol, or propagation characteristics of the environment.
Therefore, this approach can be implemented independently of specific wireless standards, and covers a range of different applications, such as localizing interfering nodes within a private wireless network.


Differently from the single transmitter case, where the \ac{WSN} collects \ac{RSS} measurements coming from a single source, the \ac{RSS} obtained at each \ac{SU} when multiple transmitters are active is a sum of the transmitted power from different sources.
Under this scenario, and assuming the often used log-normal \ac{PLM} given by \eqref{received_power}, the \ac{RSS} at each \ac{SU} has the distribution of a sum of log-normals.
This random variable does not have a closed-form density function \cite{sum_lognormals}.
Hence, model-driven approaches rely on approximations of this distribution for deriving position estimators, e.g., \cite{quasi_em_multi_tx_loc}. 
Moreover, hardware dependent nonlinearities that affect the \ac{RSS} measurement are not captured by \eqref{received_power}.
An analysis of hardware related influences on \ac{RSS} measurement has been presented by A. Zannela in \cite{best_pratice} together with guidelines for dealing with the shortcomings stemming from these nonlinearities and propagation effects.
As the number of available \acp{SU} largely influences the accuracy of the positioning algorithms, it is reasonable to assume that inexpensive \acp{SU} with limited hardware capabilities are preferred for implementing a dense \ac{WSN}. 
Consequently, the solutions of the \ac{MTL} problem that present high localization accuracy also present a high degree of implementation complexity due to the lack of tractable mathematical models.
This motivates the adoption of \ac{DL} as an approach for addressing this problem, since underling hardware-induced nonlinear patterns can be identified from the data collected by the \acp{SU}. 
Similar works on RF-based transmitter identification have shown promising results \cite{DL_user_detec,Spectrum_Occupancy_Prediction}.

More recent and closely related works on \ac{DL}-based \ac{MTL} often seek to represent position related information, i.e., \ac{RSS}, \ac{CSI}, \ac{ToA} or \ac{AoA}, in formats that resemble the structure of images, so aiming at taking advantage of the well-developed \ac{CNN} architectures employed in object recognition tasks \cite{deeptxfinder,deepmtl,dl_loc_lim_data}.
In contrast, we argue that a simpler, and consequently less computationally expensive, \ac{DNN} architecture already suffices to achieve good generalization performance for the \ac{MTL} task. 
In previous works, we proposed a \ac{DNN} architecture using solely fully connected nodes, and investigated its performance when compared to classical and state-of-the-art approaches \cite{blind_tx_loc}.
In \cite{experimental_loc}, we employed the proposed scheme to analyze its localization performance when real-world \ac{RSS} measurements \cite{dataset} are available, and to understand the achievable real-time performance by employing such technique. 
The research carried out in previous works now motivates the investigation on the scalability of the \ac{DL} approach for localization of multiple simultaneous active transmitters.

The proposed approach is divided into two stages.
Firstly, the number of active transmitters is estimated using a classification \ac{DNN} model, and secondly the corresponding transmitters' coordinates are estimated by a regression \ac{DNN} model trained for a specific number of active transmitters.
Both tasks rely solely on \ac{RSS} measurements collected by a \ac{WSN} as input information.
The proposed architecture employed for positioning is also used for estimating the number of active transmitters present in the area of interest.

In this work, the proposed approach is compared against two other approaches with different degrees of computational complexity.
Namely, (\textit{i}) \ac{REML} \cite{reml}, where an interpolated \ac{RSS} map of arbitrary pixel resolution is constructed with the measurements collected by the \ac{WSN}. 
In the case of a single active transmitter, the position of the active transmitters is then estimated as the pixel coordinate that contains the highest \ac{RSS} value, whereas for \ac{MTL}, high power regions are first identified, and the corresponding transmitters' coordinates are estimated as the center of such regions; (\textit{ii}) \ac{PS}, where the localization problem is modeled analogously to a physical particle simulation \cite{particle_simulation}.

The remainder of the paper is organized as follows: 
Section \ref{sec:sec_02} presents the mathematical model that characterizes the propagation phenomena.
Section \ref{sec:sec_03} describes the localization algorithms.
Section \ref{sec:sec_04} presents an analysis of the performance achieved by the localization schemes. 
Finally, the paper is concluded in Section \ref{sec:conclusion}.
\section{System Model}\label{sec:sec_02}
\begin{figure}[t]
	\centering
	\includegraphics[width=0.85\columnwidth]{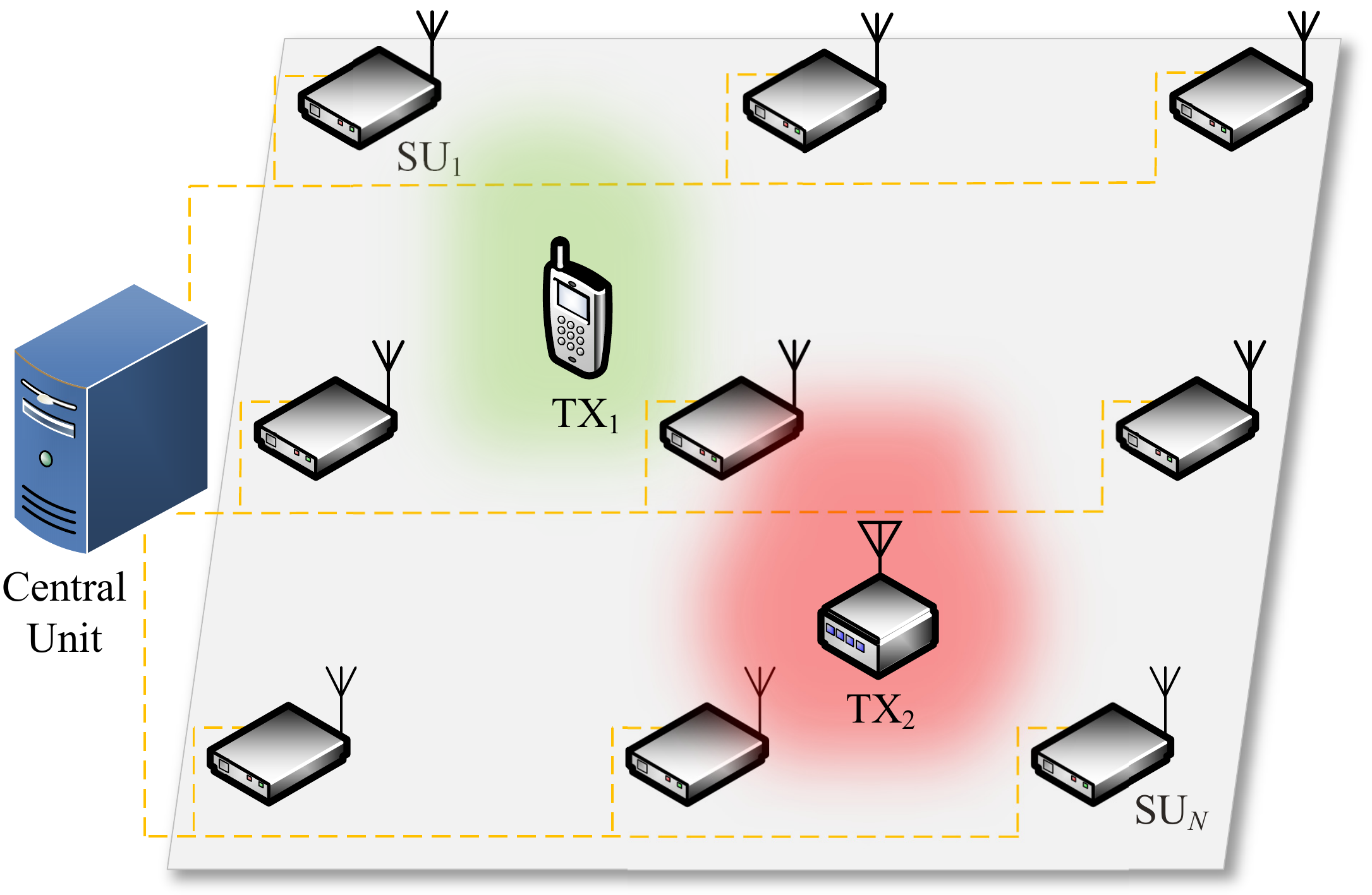}
	\caption{Illustration of a wireless network hosting the \ac{DL}-based localization service. The \ac{RSS} measurements are collected by the \acp{SU}, which are spatially distributed in grid, and sent to the \ac{CU}, where the number and the relative coordinates of the active transmitters are estimated using a selection \ac{DNN} models.}
	\label{wsn}
\end{figure}

As illustrated in Fig.~\ref{wsn}, let us consider that within the area of interest, $N_s$ \acp{SU} are placed in fixed locations and connected to a \ac{CU} forming a \ac{WSN}. 
The \acp{SU} collect \ac{RSS} measurements in a frequency band for sufficient time to average out small scale fading effects, and send them to a \ac{CU}, where the number of active transmitters and their corresponding coordinates are estimated.

Let $\mathbf{p} = \left[P_{1}, \, \cdots, \, P_{N_s}\right]^{\mathrm{T}}$ represent the measurement vector, which contains the \ac{RSS} values measured by $N_s$ \acp{SU}, where $P_j = 10\log_{10}\left(p_j\right)$ dBW, and the $j$-th \ac{SU} measurement in linear scale can be expressed as
\begin{equation} \label{received_power}
	p_j = \sum_{i=1}^{N_t} p_{0i} \left(\frac{d\left(\mathbf{u}_i, \mathbf{v}_j\right)}{d_0} \right)^{-\beta} w_{ij},
\end{equation}
where $N_t$ is the total number of active transmitters, $p_{0i}$ is the received power at a reference distance $d_0$ from the $i$-th transmitter, $\beta$ represents the path loss exponent, which depends on the environment.
$\mathbf{u}_i \triangleq \left[u_{i_x}, u_{i_y}\right]^{\mathrm{T}}$ contains the unknown coordinates of the $i$-th transmitter in two dimensions, however, extension to three dimensions is straightforward, $\mathbf{v}_j \triangleq \left[v_{j_x}, v_{j_y}\right]^{\mathrm{T}}$ are the coordinates of the $j$-th \ac{SU}, and $d\left(\mathbf{u}_i, \mathbf{v}_j\right)$ denotes the Euclidean distance between the $i$-th transmitter and $j$-th \ac{SU}. 
Lastly, $w_{ij} = 10^{n_{ij}/10}$ accounts for the random power fluctuations due to multi-path propagation and random movement, i.e., shadowing noise, where $n_{ij}$ are the entries from a zero-mean Gaussian random vector $\mathbf{n}_i$ with covariance matrix $\mathbf{C} \in \mathbb{R}^{N_s \times N_s}$.
The spatial correlation of the shadowing noise is modeled with an exponential decay.
Therefore, the entries of $\mathbf{C}$ are a function of the distances among \acp{SU} and the decorrelation distance, which is assumed to be in the order of 1 meter for indoor propagation  \cite{decorrelation_distance}.


\section{Localization Approaches}\label{sec:sec_03}

\subsection{Deep Learning Based Localization}
\begin{figure}[t!]
	\centering
	\includegraphics[width=0.9\columnwidth]{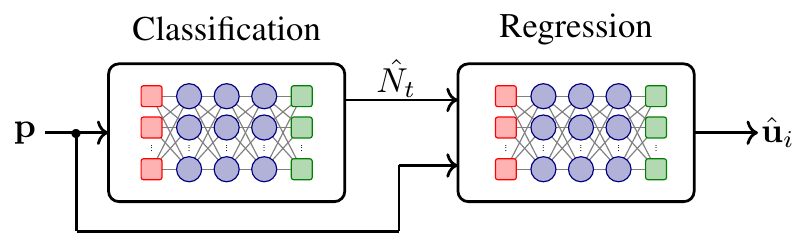}
	\caption{Diagram of the two stage \ac{DL}-based localization scheme.}
	\label{detection_localization}
\end{figure}

Our approach is divided  into two steps, and it can be classified as non-cooperative as the transmitters can be simultaneously active.
Firstly, the number of active transmitters is estimated.
This task can be carried out by different techniques, such as energy detection and cyclostationary feature detection \cite{eg1,eg2,eg3}.
However, the proposed \ac{DNN} architecture can also be trained for this classification task with a modification of its last layer, and loss function.
Secondly, the estimation of the coordinates is carried out by the \ac{DNN} architecture selected based on the outcome from the first step, i.e., a particular architecture is used depending on the number of active transmitters to be localized.
Note that the only architecture modification needed is a different number of output units, e.g., for localizing two transmitters, the architecture requires four output units for two dimensional positioning, and six units for three dimensions.
Both tasks can be modeled as supervised learning problems, given a data set with transmitters coordinates, or number of active transmitters, and associated \ac{RSS} measurements.
The approach is illustrated in Fig.~\ref{detection_localization}.

The \ac{DNN} architecture employed in this paper has been proposed in our previous work \cite{blind_tx_loc}, where we draw inspiration from the well-investigated log-normal \ac{PLM} that associates \ac{RSS} to the transmitter-receiver separation distance as a starting point for selecting the architecture.
This led to a fully connected \ac{DNN}, where the number of hidden layers is chosen based on the number of nonlinear functions between the transmitter position and the corresponding \ac{RSS} measurements.
Observing \eqref{received_power}, there is a two-fold nonlinear functional relation between the transmitters coordinates and \ac{RSS} measurements. 
Moreover, taking into account the nonlinear effects induced by the hardware, an architecture with 3 hidden layers was selected.
It is worth noting that, a single hidden layer architecture is capable of approximating any function given a sufficiently large number of hidden units \cite{universal_approximation}.
However, successful practical examples of \ac{DL} algorithms suggest that architectures with multiple hidden layers are able to approximate complex functions with significant less hidden units \cite{deep_learning}.

Differently from \cite{blind_tx_loc}, the \ac{ELU} is chosen as activation function instead of the \ac{ReLU}, since its output does not produces zero for negative inputs, and thus deactivating neurons. Furthermore, its gradient is continuous, and according to our most recent experiments \ac{ELU} performs slightly better than \ac{ReLU} w.r.t. localization accuracy.
The hyper-parameters of the selected architecture are presented in Table \ref{table_hyperparameters}.
\begin{table}[t]
	\centering
	\caption{\ac{DNN} hyperparameters.}
	\label{table_hyperparameters}
	\renewcommand{\arraystretch}{1.2}
	\begin{tabular}{ll}
		\toprule[0.9pt] 
		Hyperarameter & Value \\ 
		\midrule
		Number of hidden layers & 3 \\
		Number of hidden units per layer & 128 \\
		Validation split & 80\% training, 20\% validation \\
		Mini batch size & 40 \\
		Regularization parameter (L2) & 0.01 \\ 
		Activation function of hidden units & \ac{ELU} \\ 	
		Optimizer & Adaptive moments (Adam) \\ 
		Learning rate & 10\textsuperscript{-4} \\
		Loss function & \ac{MSE} \\
		Weight initialization & Xavier \\
		\bottomrule[0.9pt]	
	\end{tabular}
\end{table}
For obtaining the transmitters' coordinates, the last layer has a linear activation function, and its number of units depends upon the number of simultaneous active transmitters, i.e., twice the number of transmitters for two dimensional positioning.

An ambiguity problem arises depending on how the data set for training is organized. 
To illustrate this problem, let us assume the presence of two transmitters, the same measurement vector can be obtained when $\mathbf{u} = \left[u_{1_x}, u_{1_y}, u_{2_x}, u_{2_y} \right]$ and when $\mathbf{u} = \left[u_{2_x}, u_{2_y}, u_{1_x}, u_{1_y} \right]$.
To avoid such problem in the data set creation, the transmitter coordinates are ordered such that transmitters with smaller indices have the smaller coordinate values for a given example.
In other words, the known coordinates are arranged such that $u_{i_x}, u_{i_y} < u_{(i+1)_x}, u_{(i+1)_y}$.
Note that this does not pose a constraint on the functionality of the proposed solution, since distinguishing individual transmitters' indexes might not be of interest for practical applications, such as unauthorized transmitter localization.

The online computational complexity is represented by the number of real multiplications required by the \ac{DNN} estimator, and it can be written w.r.t. the network architecture as
\begin{equation} \label{complexity_dnn}
\mathcal{C}_{\mathrm{DNN}} = \sum_{l=1}^{L-1} \left(N_u^{(l-1)} + 1\right) N_u^{(l)},
\end{equation}
where $N_u^{(l)}$ represents the number of units in the $l$-th layer of the network, $L$ is the total number of layers including input and output layers.

\subsection{Radio Environment Map Localization (REML)}

This localization approach divides the area of interest into discrete regions with resolution $R$ meters, and it uses an estimated REM for obtaining the transmitters coordinates \cite{reml}.
The REM is acquired via ordinary Kriging interpolation, where the \ac{RSS} predictions at arbitrary pixel points are estimated based on the second order statistics of the measured \ac{RSS} \cite{kriging_interpolation}.
The resulting REM contains the estimated RSS values for all pixels in the area using \ac{RSS} measurements for the pixels that contain the \acp{SU}. 

Let $P_w = A_w/R$ and $P_h = A_h/R$ represent the total number of pixels along the width and height of the area, where $A_w$ and $A_h$ are the width and height in meters, respectively.
The total number of pixels within the area is $K = P_w \times P_h$ pixels.
For avoiding \ac{RSS} variation within one pixel, $R$ has to be greater than the decorrelation distance.
The REM is then stored in a matrix $\mathbf{U} \in \mathbb{R}^{P_w \times P_h}$.
Assuming a single transmitter, the pixel coordinates are estimated by selecting the element which contains the highest \ac{RSS} value as
\begin{equation}
	\mathbf{\hat{u}}_{\mathrm{pixel}} = \underset{i,j}{\arg\max}\left[\mathbf{U}\right]_{i,j},
\end{equation}
where the continuous space coordinates are obtained by $\mathbf{\hat{u}}_{\mathrm{REML}} = \mathbf{\hat{u}}_{\mathrm{pixel}}/R$.
For the case of multiple transmitters, histogram thresholding and image segmentation need to be applied to $\mathbf{U}$ for distinguishing regions with significant \ac{RSS}.
After this procedure, the transmitters pixel coordinates are estimated as the pixel with highest \ac{RSS} within each region.
It is worth noting that \ac{REML} does not require previous knowledge on the transmit power or number of active transmitters, making it also practical for blind localization.
The \ac{REML} computational complexity is given by 
\begin{equation}
	\mathcal{C}_{\mathrm{REML}} = KN_s\log(N_s).
\end{equation}

\subsection{Particle Simulation}

This localization method is based on a particle simulation. 
Therefore, all \acp{SU} and active transmitters are assumed to be particles. 
The \acp{SU} are fixed and initialized at their corresponding known positions. 
In contrast, the transmitters are initialized close to the \acp{SU} with the strongest \ac{RSS} measurements, and they can move freely during the algorithm iterations. 
Initially, the \ac{RSS} will be calculated for each \ac{SU} based on their current positions and the \ac{PLM}. 
Since the initial transmitter positions differ from the true positions, and there are other influences such as shadowing, the calculated \ac{RSS} will differ from the measured \ac{RSS} at the \acp{SU}. 
These errors are interpreted as potentials that will imply some forces onto the particles. 
Hence, the \acp{SU} that have measured more than calculated become attractors, whereas the others will become repellers. 
The forces induce movement on the free particles, i.e., transmitters, such that they will iteratively move closer to the best position that matches the measured \ac{RSS}. 
Note that in this scheme, the transmit power is assumed to be known in advance.
More details may be found in \cite{particle_simulation}.

The algorithm iterations stop, if one of the following conditions is met: 
(\textit{i}) the number of iterations exceeds $N_{\mathrm{iter}}=500$, 
(\textit{ii}) the total movement of the particles is smaller than 10\textsuperscript{-3} meters, or 
(\textit{iii}) the Euclidian norm of the vector of errors in receive powers is smaller than 10\textsuperscript{-2} dB. 
The computational complexity is dominated by the \ac{PLM} calculation for each pair of transmitter and \ac{SU} that is required to obtain the forces that affect particles in each iteration. 
Hence, the computational complexity is given by the product 
\begin{equation}
	\mathcal{C}_{\mathrm{PS}} = N_sN_tN_{\mathrm{iter}}
\end{equation}

\section{Simulation Results Regarding Scalability}\label{sec:sec_04}

The performance shown in this section focuses on the scalability of the \ac{DL} approach with respect to sensor density, and number of simultaneous active transmitters.
As localization performance metric, we present the \ac{CDF} of the localization error, where this error is defined as the Euclidean distance between the true position and the estimated one. 
For comparison we tested the data with the approaches in \cite{particle_simulation}, and in \cite{reml}. 
For the sake of comparison with the more sophisticated techniques, the performance obtained with \ac{RG} is also presented as an upper bound for the positioning error.
In this case, the transmitters' coordinates are estimated by drawing realizations from a uniform distribution ranging between 0 and the area limits. 

The data sets used are obtained following the model described by \eqref{received_power}.
The path loss exponent is fixed to $\beta = 3.23$, which corresponds to an indoor space with relatively high signal attenuation \cite{PLE_experimental_analysis}. 
A maximum of four simultaneous transmitters coexist in the area. 
This quantity is assumed to be known a priori for Fig.~\ref{cdf_loc_err_multi_tx}, and it is used for selecting the required \ac{DNN} architecture for localization.
Therefore, one \ac{DNN} model is trained for each $N_t$.
The coordinates of the transmitters at each example are generated through an uniform distribution between zero and the area limits.
The shadowing noise variance of the training and test data sets is fixed and equal to 10 dB.
Note that for training and testing, independent data sets were generated.
All relevant simulation parameters are shown in Table \ref{sim_par}.
\begin{table}[t!]
	\centering
	\caption{Simulation parameters.}
	\label{sim_par}
	\renewcommand{\arraystretch}{1.2}
	\begin{tabular}{ll}
		\toprule[0.9pt] 
		Parameter & Value \\ 
		\midrule
		Fixed transmit power & 20 dBm \\
		Antenna radiation pattern & Omnidirectional \\
		Path loss exponent ($\beta$) & 3.23 \\
		Shadowing noise variance & $\sigma^2_{\mathrm{dB}} = 10$ \\
		Decorrelation distance & 1 meter \\
		Area size & 20 $\times$ 20 m\textsuperscript{2} \\
		Number of \acp{SU} & 16 (Figs.~\ref{confusion_matrix}, \ref{cdf_loc_err_multi_tx}, \ref{rmse_matrix}) \\ 
		Sensor density & 4 sensor/100 m\textsuperscript{2} (Figs.~\ref{confusion_matrix}, \ref{cdf_loc_err_multi_tx}, \ref{rmse_matrix}) \\
		Number of active transmitters ($N_t$) & 1, 2, 3 and 4 \\ 
		\ac{SU} arrangement & Grid \\	
		Frequency of operation & 2.4 GHz \\ 
		\ac{REML} pixel resolution & 10 cm \\
		\ac{PS} maximum number of iterations & 500 \\
		Number of training examples & 3000 per active transmitter \\
		Number of testing examples & 3000 per active transmitter \\
		Number of training epochs & 1000 \\
		\bottomrule[0.9pt]	
	\end{tabular}
\end{table}

\subsection{Multiple Active Transmitters}
\subsubsection{Classification Performance}
\begin{figure}[t!]
	\centering
	\includegraphics[]{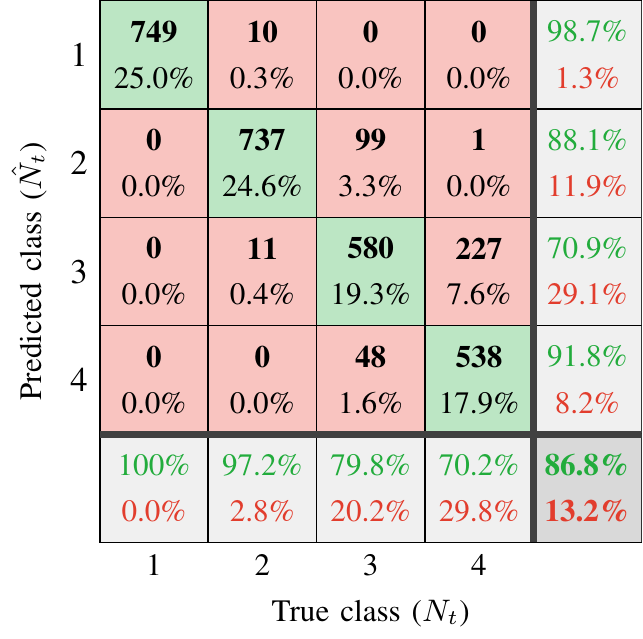}
	\caption{Confusion matrix.}
	\label{confusion_matrix}
\end{figure}
Fig.~\ref{confusion_matrix} illustrates the classification performance of the \ac{DNN} model assuming up to four simultaneously active transmitters.
The rightmost column of the confusion matrix displays the precision of the model, whereas the row in the bottom presents the recall values, and the element at the bottom right displays the model's overall accuracy.
As the number of active transmitters increases, the classification performance decreases. 
Nevertheless, precision and recall values do not fall below 70\% for the total of 3000 examples tested.

\subsubsection{Localization Performance}
\begin{figure}[t!]
	\centering
	\includegraphics[width=0.9\columnwidth]{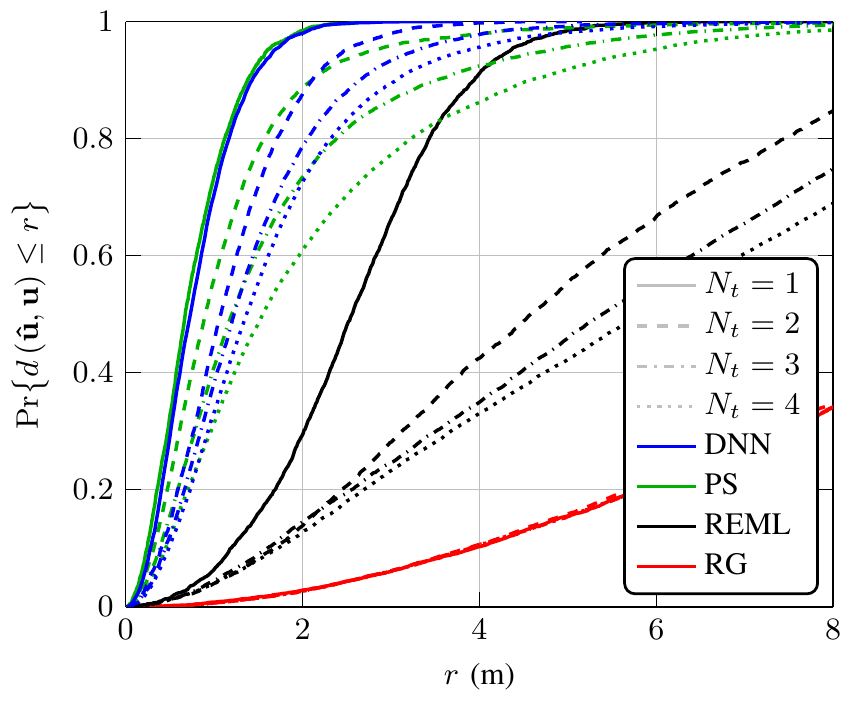}
	\caption{\ac{CDF} of the positioning error with a varying number of simultaneously active transmitters for the \ac{DL}, \ac{PS} and \ac{REML} algorithms.}
	\label{cdf_loc_err_multi_tx}
\end{figure}
Fig.~\ref{cdf_loc_err_multi_tx} shows the localization performance of the algorithms assuming up to four simultaneous transmitters.
The \ac{DL} approach yields sub 4 meter accuracy for all data tested, with performance degradation as the number of active transmitters increases.
In Fig.~\ref{cdf_loc_err_multi_tx}, one \ac{DNN} model was obtained for each number of simultaneously active transmitters.
The \ac{REML} shows acceptable performance only for $N_t<2$. 
For $N_t \geq 2$, the likelihood that two transmitters are close enough such that the regions of high \ac{RSS} levels cannot be distinguished is significant, and the localization accuracy reduces for this reason.
The \ac{PS} method shows similar performance when compared to \ac{DL} for $N_t \leq 2$.
However, \ac{PS} presents slight localization performance degradation as the number of transmitters increases.

\begin{figure}[t!]
	\centering
	\includegraphics[]{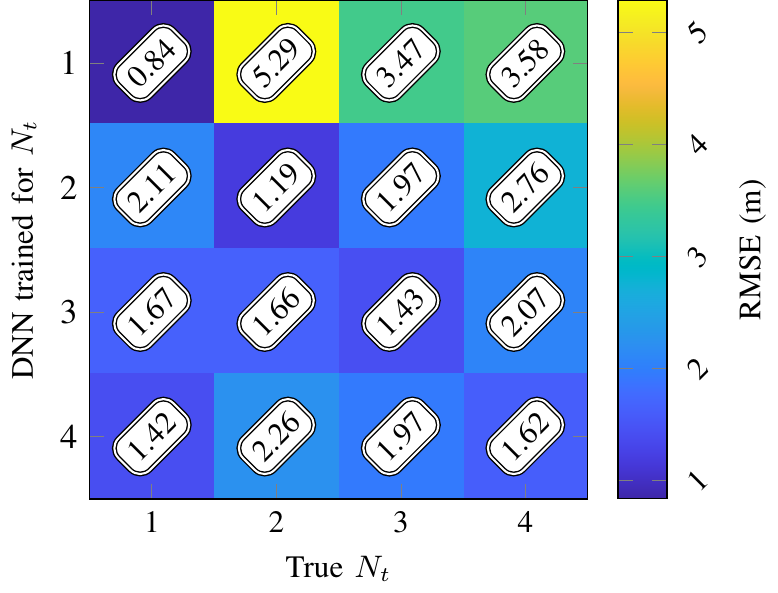}
	\caption{\Ac{RMSE} matrix of the estimated coordinates when $N_t$ is misclassified.}
	\label{rmse_matrix}
\end{figure}
Fig.~\ref{rmse_matrix} illustrates the \ac{RMSE} matrix of the estimated coordinates when $N_t$ is misclassified.
The \ac{RMSE} is obtained by using the closest coordinates when $N_t$ is misclassified.
As it can be observed the largest positioning errors occur when there are more active transmitters than what was obtained in the classification stage, i.e, a model trained for a larger $N_t$ performs better than a model trained for smaller $N_t$ in a misclassification scenario.
This result resonates with the idea that a \ac{DNN} model can only perform well when it is exposed to data that it is similar to the data contained within the training set. 
Correspondingly, the model trained for $N_t=4$ performs better than the model trained for $N_t=1$ when $N_t$ is misclassified.
Furthermore, a joint analysis of Figs.~\ref{confusion_matrix} and \ref{rmse_matrix} shows that when misclassification errors are more likely, i.e., larger $N_t$, the average positioning error is less degraded.

\subsection{Sensor Density}
\begin{figure}[t!]
	\centering
	\includegraphics[width=0.9\columnwidth]{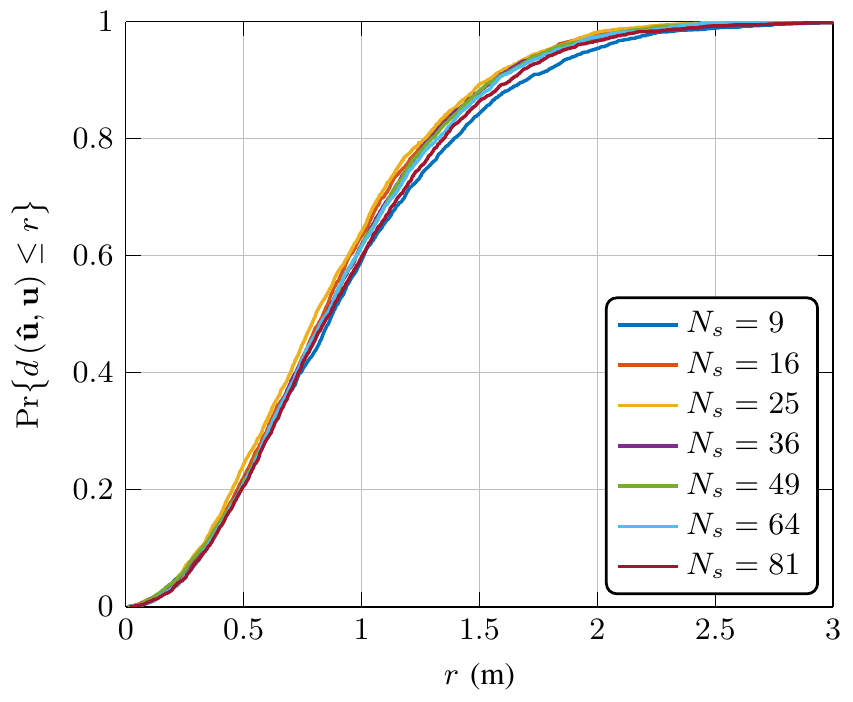}
	\caption{\ac{CDF} of the positioning error assuming a constant sensor density of 4 sensor/100 m\textsuperscript{2} and $N_t=1$ for the \ac{DL} approach.}
	\label{cdf_loc_err_su_density}
\end{figure}
Fig.~\ref{cdf_loc_err_su_density} shows the localization performance of the \ac{DL} approach for the single transmitter case assuming a constant sensor density, i.e., as $N_s$ increases the area size increases proportionally in order to keep the ratio $(A_w \times A_h)/N_s$ constant.
As can be observed, the resulting localization performance remains also constant.
This result indicates that the localization performance observed in a specific scenario, i.e., area size and $N_s$, can be extrapolated to a more general setting.

\begin{figure}[t!]
	\centering
	\includegraphics[width=0.9\columnwidth]{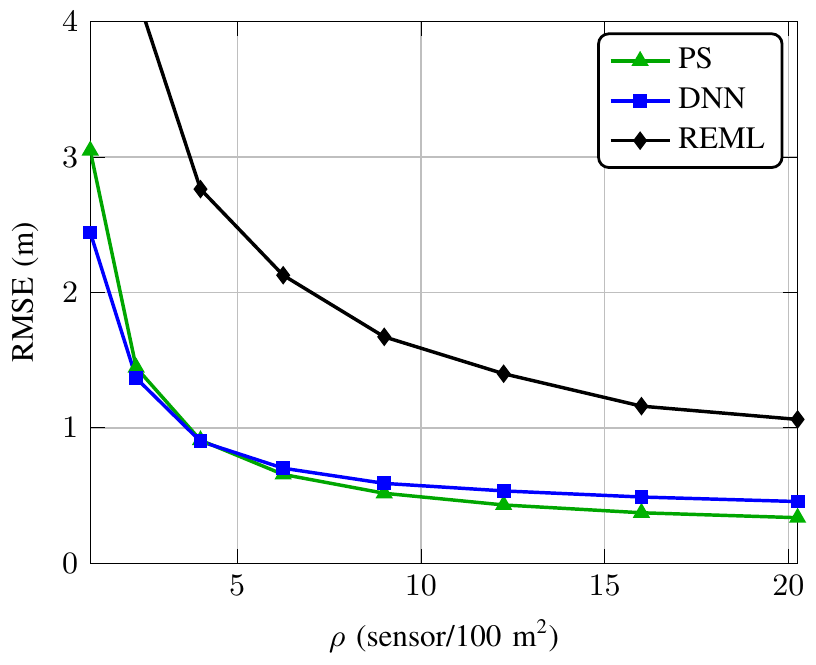}
	\caption{\Ac{RMSE} of the estimated coordinates versus \ac{SU} density assuming a constant area of 20 by 20 m\textsuperscript{2} and $N_t=1$ for the \ac{DL}, \ac{PS} and \ac{REML} approaches.}
	\label{av_rmse_vs_density}
\end{figure}
Fig.~\ref{av_rmse_vs_density} presents the localization accuracy averaged over several positions across the area as a function of the \ac{SU} density in sensor/100 m\textsuperscript{2}.
Unsurprisingly, lower \ac{SU} density leads to lower localization accuracy for all localization algorithms.
Moreover, in the lower \ac{SU} density \ac{DNN} performs slightly better than \ac{PS}, this results can be attributed to the limited number of iterations used in the particle simulation.
In the higher \ac{SU} density regime, \ac{PS} outperforms \ac{DNN} also by a small margin, since \ac{PS} requires the extra apriori information about the transmit power.
Furthermore, for all algorithms the gain in accuracy is relatively low for $\rho>15$ sensor/100 m\textsuperscript{2}, which when compared to the cost increase in \acp{SU} installation, suggests that no significant improvement on localization performance can be obtained at the expense of increasing \ac{SU} density.

\section{Conclusion}\label{sec:conclusion}

In this paper, the performance of a low complexity \ac{DL}-based localization framework has been investigated for positioning multiple transmitters in indoor scenarios.
The observed simulation results indicate that this approach is able to address the challenging \ac{MTL} problem, and its localization performance scales well with an increasing number of active transmitters.
Furthermore, it has been observed that evaluating the localization performance with respect to \ac{SU} density in scaling environments with similar characteristics is sufficient to gather insights on the expected positioning accuracy.
Nevertheless, the localization performance saturates after a certain \ac{SU} density. 
This observation suggests that more input information rather than exclusively \ac{RSS} measurements is required for achieving sub-meter localization accuracies. 
\section*{Acknowledgment}
This work was supported by the European Union's Horizon 2020 research and innovation programme through the project "iNGENIOUS" under grant agreement 957216, by the German Research Foundation (DFG, Deutsche Forschungsgemeinschaft) as part of Germany's Excellence Strategy - EXC 2050/1 - Project ID 390696704 - Cluster of Excellence "Centre for Tactile Internet with Human-in-the-Loop" (CeTI) and by the German Federal Ministry of Education and Research (BMBF) through the projects "Industrial Radio Lab Germany (IRLG)" under contract 16KIS1010K and "6G-life" under contract 16KISK001K. We also thank the Center for Information Services and High Performance Computing (ZIH) at TU Dresden.
\bibliographystyle{ieeetr}
\bibliography{my_references}
\end{document}